\documentstyle[preprint,aps,tabularx]{revtex}

\begin{document}

\title{Family Unification, Exotic States and Magnetic Monopoles}

\author{{\bf  Thomas W. Kephart$^{(a)}$ and Qaisar Shafi$^{(b)}$}}
\address{(a)Department of Physics and Astronomy,\\
Vanderbilt University, Nashville, TN 37325.}
\address{(b)Bartol Research Institute,\\
University of Delaware, Newark, DE 19716.}

\maketitle

\begin{abstract}
The embedding in $SU(4)\times SU(3)\times SU(3)$
of the well studied gauge groups $SU(4)\times SU(2)\times SU(2)$ and
$SU(3)\times SU(3)\times SU(3)$ naturally leads to family unification as opposed
to simple family replication.
An inescapable consequence is the predicted existence of (exotic)
color singlet states that carry fractional electric charge.
The corresponding magnetic monopoles carry multiple Dirac 
magnetic charge, can be relatively light $(\sim 10^{7}-10^{13}GeV)$,
and may be present in the galaxy not far below the Parker bound.
\end{abstract}



\newpage

\

\bigskip

\bigskip
Much work has been done in trying to extract standard model physics from
string theory, beginning with Calabi-Yau compactifications of the
heterotic
string, which yield $E_{6}$ type $GUT$ theories, where holomorphic
deformations, Wilson loops, etc. can be used to reduce the gauge
symmetry,
and continuing today with orbifolding of  type $IIB$ strings on $%
AdS_{5}\otimes S^{5}$ to produce conformal field theories ($CFT$s$)$
with
gauge groups $\prod_{i}SU(Nd_{i})$ and bifundamental matter
\cite{Lawrence:1998ja},
\cite{Kachru:1998ys}.
Here we
take a
bottom-up approach and consider a model that is likely to be deriveable
from
orbifolded type $IIB$ strings, but focus more on the physics that will
result. The model we study contains aspects of both $E_{6}$ and $CFT$
type
string theory compactifications, and leads to a remarkably rich
phenomenology.
It is well known that the Pati-Salam (PS) model
\cite{Pati:1974yy}
and the Trinification
\cite{trin}
(TR)
model are both contained in $E_{6}$ Grand Unification
\cite{Gursey:1976ki},
\cite{Achiman}. 
We will provide
another covering of PS and TR which does not embed in $E_{6}$, but
instead
{\it requires} 3 families and is perhaps the minimal such example of a model
with
these properties. After a brief review of PS and TR, we present our model
and
then consider some of its consequences.

The PS model has gauge group $G_{PS}=SU(4)\times SU_{L}(2)\times
SU_{R}(2)$,
and fermion families in bifundamental representations
\begin{equation}
(4,2,1)+(\bar{4},1,2)
\end{equation}
This model embeds in $SO(10)$, where the fermions are then all contained
in
a {\bf16}. Adding a {\bf10} + {\bf1} of fermions then allows unification into $E_{6}$,
where the fermions are then in a {\bf27}. Various aproaches to symmetry
breaking
have been studied, but we need not be concerned with these details
for
now. Note that this model is anomaly free for a single family, so a full
three-family model is gotten simply by the inclusion of two more
families.
TR also has fermions in bifundamental representations of the gauge group
$G_{TR}=SU(3)\times SU(3)\times SU(3):$
\begin{equation}
(3,\bar{3},1)+(\bar{3},1,3)+(1,3,\bar{3})
\end{equation}
As this group is already a maximal subgroup of $E_{6}$ and (2) already
contains 27 states, the unification into $E_{6}$ is gotten simply by
adding
the necessary gauge generators to extend $SU^3(3)$ to $E_{6}$. Again,
a
single fermion family is anomaly free on its own, so we must add two
families to agree with phenomenology.

The smallest group that contains
both
the SP model and TR is not $E_{6}$ but $G=SU(4)\times SU(3)\times
SU(3)$,
which has 31 generators and is rank 7. Insisting on fermions that fall
into
bifundamentals, the representations to consider are $(4,\bar{3},1)$,
$(\bar{4}%
,1,3)$, and $(1,3,\bar{3})$. We can not take one of each to form a
family,
since this would be anomalous. The minimal anomaly free choice is

\begin{equation}
3(4,\bar{3},1)+3(\bar{4},1,3)+4(1,3,\bar{3})
\end{equation}

If we break $SU(4)$ to $SU(3)$, then (3) becomes

\begin{equation}
3[(3,\bar{3},1)+3(\bar{3},1,3)+3(1,3,\bar{3})]+(1,\bar{3},1)+(1,1,3)+(1,3,\bar{3})
\end{equation}
which contains three TR families plus a few additional particles. Hence the
simplest anomaly free chiral bilinear representation $[i.e.,(3)]$
contains
three families ($i. e.$, this is a true family unification) instead of one as for either PS(1), TR(2), or the {\bf27} of
$E_{6} $ (where the second and third families are gotten from merely
replicating the first). To further analyze this $SU(4)\times SU(3)\times SU(3)$ 
model\cite{footnote}
(334-model), we must consider the spontaneous symmetry breaking (SSB), and the 
charge assignments this leads
to,
plus the implication of the ``extra" fermions. We will find there exist
fractional charged color singlets
\cite{Kim:1980yk},  
\cite{Goldberg:1981jt} in the model, and 
hence the minimal
monopole change in the model will be the inverse of the minimal fraction
times the Dirac charge.

The possibilities for embedding color and weak isospin of the standard model gauge group in $%
SU(4)\times SU(3)\times SU(3)$ are:

$(i)$ Embed $SU_{C}(3)$ in one $SU(3)$ and $SU_{W}(2)$ in the other
$SU(3)$.

$(ii)$ Embed $SU_{C}(3)$ in an $SU(3)$ and $SU_{W}(2)$ in $SU(4)$.

$(iii)$ Embed $SU_{C}(3)$ in the $SU(4)$ and $SU_{W}(2)$ in an $SU(3)$.

Other embeddings are equivalent except for embedding $SU_{C}(3)$ and/or
$%
SU_{W}(2)$ in some diagonal subgroup within the 334-model. However,
this
leads to vectorlike fermions, and we need not pursue this possibility
any
further.

The embedding of weak hypercharge is more complicated. Consider the
breaking $%
SU(4)\times SU_{L}(3)\times SU_{R}(3)\rightarrow SU(4)\times
[SU_{L}(2)\times U_{A}(1)]\times [SU_{R}(2)\times U_{B}(1)]$. If we then
break $U_{A}(1)$ and $U_{B}(1)$ completely, the hypercharge must be $%
Y=T_{3R}+(B-L)$, where $T_{3R}$ is the diagonal generator of
$SU_{R}(2)$,
and $B-L$ generates the $U(1)$ that is in $SU(4)$ but not in
$SU_{C}(3)$.
However, there are other possibilities for the embedding of $U_{Y}(1)$.
These are similar, and in some cases equivalent, to the well-known
flipped
models
\cite{DeRujula:1980qc},
\cite{Cleaver:2001sc}.
One obvious choice is to break $SU(4)$ to $SU_{C}(3)$ and then
$Y$
could be the trinification choice from $SU_{L}(3)\times SU_{R}(3)$.
Similarly, trinification has a standard hypercharge assignment, but
this
could be flipped to a Pati-Salam charge assignment. Also moving
$SU_{W}(2)$
from $SU(4)$ to an $SU(3)$ of the 334-model corresponds to an isoflipped
model
\cite{Kephart:1989az}. (There are even more choices, but they will be described
elsewhere.)
Here we restrict ourselves to the standard hypercharge embeddings, but keep in mind
that
flipping may offer other opportunities.

Returning to the standard Pati-Salam version of the 334-model, we find
on
breaking $G$ to $G_{PS}$ the fermions become
\begin{equation}
3[(4,\bar{2},1)+(4,\bar{1},1)]+3[(\bar{4},1,2)+(\bar{4},1,1)]+4[(1,2,\bar{2})
+(1,2,1)+(1,1,\bar{2})+(1,1,1)]
\end{equation}
At this stage only the three families remain chiral, while the
extra (exotic) vectorlike fermions obtain masses from Higgs VEVs at the $G$ breaking scale. Now,
breaking
to the standard model $G_{SM}=SU_{C}(3)\times SU_{W}(2)\times U_{Y}(1)$,
we
find
\begin{eqnarray*}
&&3[(3,2)_{\frac{1}{6}}+(1,2)_{-\frac{1}{2}}+(3,1)_{\frac{1}{6}}+(1,1)_{-\frac{1}{2}}] \\
&&+3[(\bar{3},1)_{\frac{1}{3}}+(\bar{3},1)_{-\frac{2}{3}%
}+(1,1)_{1}+(1,1)_{0}+(\bar{3},1)_{-\frac{1}{6}}+(1,1)_{\frac{1}{2}}] \\
&&+4[(1,2)_{-\frac{1}{2}}+(1,2)_{\frac{1}{2}}+(1,2)_{0}+(1,1)_{-\frac{1}{2}}
+(1,1)_{\frac{1}{2}}+(1,1)_{0}.]
\end{eqnarray*}
 As expected, we are left with three standard families, plus three
right-handed neutrinos, from the three Pati-Salam families. In addition
we
have the extra states:
\begin{eqnarray*}
&&Q_E=3[(3,1)_{\frac{1}{6}}+(\bar{3},1)_{-\frac{1}{6}}]+4[(1,2)_{\frac{1}{2}}
+(1,2)_{-\frac{1}{2}}]+7[(1,1)_{\frac{1}{2}}+(1,1)_{-\frac{1}{2}}] \\
&&+4(1,2)_{0}+4(1,1)_{0}.
\end{eqnarray*}
\newline
Once color is confined, we see from $Q_E$ that electric charge is quantized in units of
$\frac{1}{2}$.
So any magnetic monopoles that exist in the model must have minimum
charge two from the Dirac quantization condition.

In the case of trinification, an interesting subtlety arises on
breaking $%
SU(4)$ to $SU(3)$. With the standard trinification charge assignments,
we
will find massless charged quarks and leptons. To avoid the resulting
conflict with phenomenology, we must add a few additional states at the
334
level. Let us see how this works. Two equivalent cases must be
considered:
(1) $SU_{C}(3)$ embedded in $SU(4)$, or (2) $SU_{C}(3)$ identified with
an $%
SU(3)$ of the 334-model. In both cases at the trinification level we
have
fermions as in (4). For case 1, we have the three standard families
plus
\begin{equation}
R_{E}=3(1,\bar{3},1)+3(1,1,3)+(1,3,\bar{3})
\end{equation}
under $SU_{C}(3)\times SU_{L}(3)\times SU_{R}(3)$. Hence all the extra
states are leptonic. Then for $SU_{L}(3)\times SU_{R}(3)\rightarrow  
SU_{L}(2)\times U_{L}(1)\times U_{R}(1)$ where we identify $U_{R}(1)$
with
the diagonal generator $Y_{R}=diag(1,1,-2)$ of $SU_{R}(3)$ and likewise
$%
U_{L}(1)$ is generated by $Y_{L}=diag(1,1,-2)$ of $SU_{L}(3)$, we can choose the hypercharge to be
$Y=\frac{1}{6}Y_{L}+\frac{1}{3%
}Y_{R}$. The families just have the standard ${\bf 27}$ of $E_{6}$
charges,
while the new leptons are
\begin{equation}
5(1,2)_{-\frac{1}{6}}+(1,2)_{\frac{5}{6}}+10(1,1)_{\frac{1}{3}%
}+5(1,1)_{-\frac{2}{3}}
\end{equation}
As these states are still chiral, the only way to give them mass would
be
with a VEV from an electrically charged Higgs. As this must obviously be
avoided, the alternative is to arrange these particles to be vectorlike
by
adding the conjugate, but anomaly free chiral multiplets
\begin{equation}
\bar{R}_{E}=3(1,3,1)+3(1,1,\bar{3})+(1,\bar{3},3)
\end{equation}
at the 334 level.
In this case, upon breaking $G\rightarrow G_{TR}$ at the scale $<\phi>$, 
the chiral families stay massless while 
the extra fermions acquire mass
terms  
of the form $h<\phi>$$\bar{R}_{E}R_{E}$ where $h$ is a typical Yukawa coupling constant.
Hence, the fractionally charged leptons
become heavy compared to the family fermions. Let us summarize the
extra vectorlike leptons. There are five doublets with electric charges
$\pm
\frac{1}{3}$ and $\mp \frac{2}{3}$ , two doublets with electric charge
$\pm
\frac{1}{3}$ and $\pm \frac{4}{3}$, ten singlets with $\pm \frac{1}{3}$
charges, and five singlets with  $\mp \frac{2}{3}$ charges. The minimal
monopole charge is three.

\bigskip

For case 2, some of the extra states will be colored. In terms of
$SU_{C}(3)%
\times SU_{L}(3)\times SU_{R}(3)$, they are
\begin{equation}
S_{E}=3(3,1,1)+3(1,1,\bar{3})+(\bar{3},1,3)
\end{equation}
The hypercharge in unchanged 
(we continue to ingore flipping and other possible charge assignments)
from case (1) (it is still $Y=\frac{1}{6}Y_{L}+\frac{1}{3%
}Y_{R}$), so we find:
\begin{equation}
S_{E}=3(3,1)_0+3(1,2)_{-{\frac{1}{3}}}+3(1,1)_{\frac{2}{3}}+(\bar{3},2)_{\frac{1}{3}}%
+(\bar{3},1)_{-{\frac{2}{3}}}
\end{equation}
Again we must add conjugate states
\begin{equation}
\bar{S}_{E}=3(\bar{3},1,1)+3(1,1,3)+(3,1,\bar{3})
\end{equation}
and this allows masses for the exotics at a higher scale than family masses.

The symmetry breaking scales of $SU(4)\times SU(2) \times SU(2)$
and $SU(3) \times SU(3) \times SU(3)$ determine the mass of the 
associated magnetic monopoles.  In this context it is important
to recall the absence of gauge boson mediated proton decay in
these models.  Proton decay can still proceed through higgs/higgsino 
exchange, as well as via higher dimension operators.
It is relatively straight forward, however, to construct 
models based on $G_{PS}$ and $G_{TR}$ such that these
processes are forbidden, as a 
consequence say of  an `accidental' baryon number symmetry.
This opens up the possiblity that $G_{PS}$ and $G_{TR}$ 
can be broken at scales considerably below the conventional
grand unification scale $M_{GUT}\sim10^{16}GeV$.
An example based on D-branes in Type I string theory was
recently provided for the $G_{PS}$ symmetry
\cite{Leontaris:2000hh}.  With the
standard embedding of $SU(3)_{C} \times SU(2) \times U(1)$,
the symmetry breaking scale of $G_{PS}$ turns out to be
$M_{PS}\sim 10^{12} - 10^{13}$ GeV, with the corresponding
string scale $\stackrel{_>}{_\sim} M_{PS}$.  
Thus, monopoles with mass $\sim 10^{13} - 10^{14}$ GeV
are expected in
this class of models.  
An even more suggestive result is the Pati-Salam type model based on
$CFT$
obtained from orbifolded type $IIB$ strings 
\cite{Frampton:2000zy},
\cite{Frampton:2000mq}.
Here the
unification is in the $100TeV$ range, where this gives $sin^{2}\theta
=.23$
and other intriguing phenomenology 
\cite{Frampton:2001xh}. 
There is no reason why
analogous considerations cannot be carried out
for the trinification scheme and, by extension, for the
gauge symmetry of special interest here $SU(4) \times SU(3) \times SU(3)$.
The multiply
charged monopoles of the theory will have mass $M\sim10^{7}GeV$, $i.e.$, in the preferred 
range of interest if they are to be candidates for
high
energy cosmic ray primaries
\cite{Kephart:1996bi},
\cite{Wick:2000yc}.
We expect
the
334-model to have a similar unification scale with resulting exotic
(fractionally charged) leptons and/or hadrons, and we expect their masses
to be
near this unification scale, so they are of interest as dark matter
candidates
\cite{Albuquerque:2000rk},
\cite{Chung:2001cb}.  

Given that monopoles of mass $\sim 10^{13} - 10^{14}$ GeV
(or perhaps even much lighter) can arise in realistic
models, it is important to ask:  Can primordial monopoles 
survive inflation?
If the underlying theory is non-supersymmetric, an 
inflationary scenario which dilutes but does not completely
wash away intermediate mass monopoles was developed in ref
\cite{Lazarides}.
Note that the D-brane scenario discussed above gives rise to 
non-supersymmetric $SU(4)_{C} \times SU(2) \times SU(2)$,
so the discussion in ref
\cite{Lazarides}  may be relevant.  The monopole
flux can be close to the Parker bound of order
$10^{-16} cm^{-2} s^{-1} sr^{-1}$. In the orbifolded scheme, the SSB scale 
 where the monopoles get their masses can be below 
the inflation scale. Hence the monopoles can exist in interesting densities (near the Parker bound)
depending on details of the SSB phase transitions.
For the supersymmetric case, dilution of monopoles can
be achieved by thermal inflation
\cite{Pana},
\cite{Lyth}
 followed by entropy production.
A scenario in which thermal inflation is associated with the
breaking of the $U(1)$ axion symmetry was recently developed in
ref \cite{Laz}.

In summary, the $SU(4) \times SU(3) \times SU(3)$ models we are
advocating provides a natural family unification while avoiding proton decay and
giving rise to both (exotic) fractionally charged
color singlets and corresponding multiply charged magnetic
monopoles
\cite{King:1998ia}
with masses that can be well below $M_{GUT}\sim10^{16}GeV$,
perhaps as light
as $\sim 10^7 GeV$,
(Note that in $SU(5)$ the lightest monopole has mass of
$\sim 10^{17}GeV$, and carries one unit of magnetic charge
\cite{Daniel:1980yz}.)
The exotic states are heavy (greater than a few $TeV$), but may be in the range 
explored by accelerators in the next decade. 

\section*{acknowledgements}

\noindent  
TK thanks the Bartol Research Institute at the 
University of Delaware for hospitality while this work was in progress. The work of TK and QS were supported in part by the US Department of Energy
under Grants No. DE-FG05-85ER-40226 and DE-FG02-91ER 40626.


\begin{references}
\bibitem{footnote}
An example of such a model obtained from orbifolding 
might be $S^{5}/\Gamma $, 
where $\Gamma$ contains 3 and 4 dimensional irreducible representations and
we
choose $N=1$ and require a nontrivial complex embedding of a 4
dimensional
representation (not necessarily irreducible) of $\Gamma$ in the $SU(4)$
$R$
symmetry of the underlying ${\cal N}=4$ SUSY $AdS_{5}\otimes S^{5}$ theory.
\end{references}
\end{document}